%% file: main.tex
\documentclass[manuscript,screen]{acmart}
\AtBeginDocument{%
  \providecommand\BibTeX{{%
    \normalfont B\kern-0.5em{\scshape i\kern-0.25em b}\kern-0.8em\TeX}}}

\copyrightyear{2021}
\acmYear{2021}
\setcopyright{acmcopyright}\acmConference[UMAP '21]{Proceedings of the 29th ACM Conference on User Modeling, Adaptation and Personalization}{June 21--25, 2021}{Utrecht, Netherlands}
\acmBooktitle{Proceedings of the 29th ACM Conference on User Modeling, Adaptation and Personalization (UMAP '21), June 21--25, 2021, Utrecht, Netherlands}
\acmPrice{15.00}
\acmDOI{10.1145/3450613.3456831}
\acmISBN{978-1-4503-8366-0/21/06}

\usepackage[linesnumbered, ruled]{algorithm2e}
\SetKwComment{Comment}{$\triangleright$\ }{}
\usepackage{enumitem}
\usepackage{amsmath}
\usepackage{fixmath}
\usepackage[font=normal,skip=0.5pt]{caption}

\begin{document}

\title{Knowledge Tracing for Complex Problem Solving: Granular Rank-Based Tensor Factorization}

%
\author{Chunpai Wang}
\email{cwang25@albany.edu}
\orcid{0000-0003-3162-4310}
\affiliation{%
  \institution{University at Albany - SUNY}
  \city{Albany}
  \state{New York}
  \country{USA}
  \postcode{12222}
}

\author{Shaghayegh Sahebi}
\email{ssahebi@albany.edu}
\affiliation{%
  \institution{University at Albany - SUNY}
  \city{Albany}
  \state{New York}
  \country{USA}
  \postcode{12222}
}

\author{Siqian Zhao}
\email{szhao2@albany.edu}
\affiliation{%
  \institution{University at Albany - SUNY}
  \city{Albany}
  \state{New York}
  \country{USA}
  \postcode{12222}
}

\author{Peter Brusilovsky}
\email{peterb@pitt.edu}
\affiliation{%
  \institution{University of Pittsburgh}
  \city{Pittsburgh}
  \state{Pennsylvania}
  \country{USA}
  \postcode{15260}
}

\author{Laura O. Moraes}
\email{laura.moraes@coppe.ufrj.br}
\affiliation{%
  \institution{Federal University of Rio de Janeiro}
  \city{Rio de Janeiro - RJ}
  \country{Brazil}
  \postcode{21941-901}
}

%
\renewcommand{\shortauthors}{Wang, et al.}

\begin{abstract}
Knowledge Tracing (KT), which aims to model student knowledge level and predict their performance, is one of the most important applications of user modeling. 
Modern KT approaches model and maintain an up-to-date state of student knowledge over a set of course concepts according to students' historical performance in attempting the problems. 
However, KT approaches were designed to model knowledge by observing relatively small problem-solving steps in Intelligent Tutoring Systems. While these approaches were applied successfully to model student knowledge by observing student solutions for simple problems, such as multiple-choice questions, they do not perform well for modeling complex problem solving in students.
Most importantly, current models assume that all problem attempts are equally valuable in quantifying current student knowledge.
However, for complex problems that involve many concepts at the same time, this assumption is deficient. 
It results in inaccurate knowledge states and unnecessary fluctuations in estimated student knowledge, especially if students guess the correct answer to a problem that they have not mastered all of its concepts or slip in answering the problem that they have already mastered all of its concepts. 
In this paper, we argue that not all attempts are equivalently important in discovering students' knowledge state, and some attempts can be summarized together to better represent student performance.
We propose a novel student knowledge tracing approach, \textit{G}ranular \textit{RA}nk based \textit{TE}nsor factorization (GRATE), that dynamically selects student attempts that can be aggregated while predicting students' performance in problems and discovering the concepts presented in them. 
Our experiments on three real-world datasets demonstrate the improved performance of GRATE, compared to the state-of-the-art baselines, in the task of student performance prediction.
Our further analysis shows that attempt aggregation eliminates the unnecessary fluctuations from students' discovered knowledge states and helps in discovering complex latent concepts in the problems.

\end{abstract}

\ccsdesc[500]{Social and professional topics~Student assessment}
\ccsdesc[500]{Computing methodologies~Tensor factorization}
\keywords{knowledge tracing, tensor factorization, complex problem solving, aggregation}


\maketitle

\section{Introduction}
\label{sec:intro}
\input{1_intro.tex}
\section{Granular Rank Based Tensor Factorization (GRATE)}
\label{sec:model}
\input{4_model.tex}
\section{Experiments}
\label{sec:experiments}
\input{5_experiments.tex}
\section{Conclusions}
\label{sec:conclusions}
\input{6_conclusions.tex}
\noindent\textbf{Acknowledgements.} This paper is based upon work supported by the National Science Foundation under Grant No. 1755910.

\bibliographystyle{ACM-Reference-Format}
\bibliography{main}

\end{document}

%% file: 1_intro.tex

Personalized online learning systems have recently drawn a lot of attention because of the growing need to assist and improve students' learning. 
A fundamental part of the user modeling task in these systems is estimating students' knowledge states as they work with learning materials~\cite{corbett1994knowledge}.
This task, known as knowledge tracing (KT), is necessary for predicting students' performance in future assessments, personalizing problems and exercises for students, identifying at-risk students, and providing teachers with a detailed view of overall student progress.
In particular, KT models use student attempt sequences, including student performance (e.g., success or failure) on past problems, to estimate student knowledge at the end of a sequence and predict student performance on the next attempts. 

To quantify student knowledge, traditional KT models rely on a predefined \textit{domain knowledge model} that represents the associations between the problems and course concepts.
Such models individually trace student knowledge in each of these concepts, neglecting the potential relationships between different concepts.
As these models learn the same set of parameters for all students, they are not personalized to the student specifications.
For example, Bayesian knowledge tracing (BKT)~\cite{corbett1994knowledge}, which is one of the pioneer KT models, represents student knowledge states in each concept using a two-state HMM, which imposes a Markovian assumption on knowledge states from one attempt to the next.

In recent years, modern KT models have been developed to address the above problems. 
For example, many variants of BKT have been proposed to improve the model by considering the potential to forget the learned concepts~\cite{hawkins2014learning}, accounting for the dependencies between concepts~\cite{kaser2017dynamic}, and personalizing the model parameters for different students~\cite{yudelson2013individualized}.  
In addition to the Bayesian models, latent factor approaches have been successful in considering the concept relationships~\cite{thai2012factorization,lan2014sparse,sahebi2016tensor,zhao2020modeling}.
For example, Lan et al.~\cite{lan2014sparse} proposed a sparse factor analysis framework for both student knowledge tracing and domain knowledge estimation. 
Sahebi et al.~\cite{sahebi2016tensor} proposed a tensor factorization method to explicitly model student learning processes by assuming a strictly monotonic increasing learning gain.
Zhao etal.~\cite{zhao2020modeling} leverage the multi\-view tensor factorization method for modeling student knowledge using multiple learning resource types.  
Similarly, deep learning models, such as DKT~\cite{piech2015deep} and DKVMN~\cite{zhang2017dynamic}, have recently been introduced into the KT domain.



However, the majority of KT models have assumed that each attempt in a sequence considered by tracing is relatively simple and involves the application of one or very few concepts, such as small steps in solving either a complex problem or an elementary problem.
With this assumption, the observed student performance can be directly associated with a few involved domain concepts, and each correct or incorrect attempt by the student can provide a relatively confident evaluation of student knowledge in those concepts.
As a result, when considering these kinds of problems, current KT models assume that every attempt in student history is equally important in quantifying student knowledge.
This assumption can be sufficient for domains in which each problem consists of a few atomic concepts.
However, it is deficient for domains with more complex problems, such as writing a program or solving an assignment with multiple steps.

In complex problem solving, each problem can include multiple concepts, such that knowing all of them to some extent is necessary for correctly answering the problem.
Because of this complexity, student attempt observations will be noisier, as slipping in even one of the required concepts can significantly harm student performance.
Additionally, identifying the concepts that are responsible for an imperfect performance will be more challenging in such complex problems.
Similarly, solving a complex problem correctly by guessing a difficult unknown concept or by trial and error on that important concept will be wrongly attributed to a student's high knowledge of all of the involved concepts.
As a result, such noisy observations could easily cause traditional KT models to provide an inaccurate estimation of overall levels of student knowledge.
For example, consider a student who has already mastered some concepts. 
This student tries a problem on those concepts three times, getting the problem right the first time (successful attempt), slipping in one of the concepts the second time (failed attempt), and getting it right again in the third time (successful attempt).
In current KT approaches, since the model tries to fit every student attempt, these cases result in fluctuations in estimated student knowledge.
Even in models like BKT, which try to consider small guess and slip probabilities by modeling each concept independently, or DKT+~\cite{yeung2018addressing}, which aims to smooth out student predicted performance (not knowledge) using a constraint, having a binary knowledge state and fitting to every attempt results in knowledge state inaccuracies.

In this paper, we argue that, due to the noise in solving complex problems, some student attempt observations are more informative and important than others.
For that, we address the student knowledge tracing challenge for complex problems by summarizing student attempts to better represent student performance.
We propose a personalized knowledge tracing model that automatically detects ``less important'' student attempts and aggregates them into other attempts to better represent student knowledge and predict their performance.
Additionally, our proposed KT model is personalized for students and automatically discovers the domain knowledge model without requiring extra problem information, such as text, topics, or tags. 
In particular, we model student sequences in a tensor and propose an adaptive \textit{G}ranular \textit{RA}nk based \textit{TE}nsor factorization (GRATE) to address the noisiness and sparsity issues so as to provide a plausible and precise knowledge modeling. 
We impose a rank-based constraint on student knowledge across attempts to help reduce the unnecessary fluctuations in student knowledge and improve the interpretability of the model.
GRATE does not rely on a domain knowledge model, as it automatically discovers latent concepts for the problems presented to students.
It is personalized by profiling students into student latent features and learning a separate set of parameters for them in a collaborative way.




Our contributions in this paper are: \textbf{(a)}, we are the first to address the noisy observation challenge for student knowledge tracing in complex problem solving; \textbf{(b)}, for this, we propose a novel tensor factorization method that adaptively aggregates student attempts while imposing a rank-based constraint to represent students' gradual learning; \textbf{(c)}, our knowledge tracing model is personalized and does not rely on additional domain knowledge information; \textbf{(d)}, we conduct extensive experiments to analyze and validate the effectiveness of our proposed model, compared to several state-of-the-art baselines, on three real-world datasets; and \textbf{(e)}, we demonstrate that our proposed method is capable of providing precise and plausible student knowledge states while learning meaningful question-concept associations.


%% file: 4_model.tex
Our goal in this work is to handle the noise and fluctuations in student knowledge tracing of complex problems without relying on a domain knowledge model or a predefined mapping between problems and concepts.
We aim to do this with a personalized KT approach that is interpretable without harming the model performance; e.g., in the student performance prediction task.
In the following, we formulate this challenge as a tensor factorization problem, present our proposed method to address the challenge, explain our intuition behind choosing tensor factorization as the basis of our model, and provide the algorithm for our method.
\subsection{Problem Formulation and Assumptions}


We consider an online learning system in which $\mathcal{M}$ students attempt $\mathcal{N}$ problems in sequences of maximum length $\mathcal{T}$ over time.
Students can attempt the problems in any order and as many times as they like.
Student performance during each attempt is recorded as a score, grade value, or binary (success or failure) data.
We represent the students' logged performance records in a 3-mode tensor $\mathbf{X}\in [0,1]^{\mathcal{M}\times \mathcal{T} \times \mathcal{N}}$. 
Every entry $x_{u,t,i} \in \mathbf{X}$ represents the $u^{th}$ student's normalized grade on $i^{th}$ problem at attempt index $t$. 
Our goal is to factorize this tensor to be able to accurately estimate student knowledge of the problems' latent concepts and predict student performance in their future problem attempts, according to their history.

\noindent\textbf{Model Assumptions.}
We build our model based on the following assumptions: 
\textbf{(a)} \textit{Domain knowledge assumption}: Each problem covers a number of concepts that are presented in the course with different proportions; the set of all of the course concepts are shared across problems; and the training data does not include the problems' contents nor their concepts.
\textbf{(b)} \textit{Student performance assumption}: Different students have different learning abilities and initial knowledge and their performance in different problems depends on their knowledge state, especially in the concepts related to those problems.
\textbf{(c)} \textit{Student learning assumption}: As students interact with the problems, they learn the concepts that are presented in them, meaning that their knowledge in these concepts increases gradually; but students may also forget some concepts.
\textbf{(d)} \textit{Attempt noisiness assumption}: Student data can be noisy; e.g., they may slip in one concept out of all the problem concepts and receive a low score, while they have already mastered all of these concepts. Similarly, they may guess the correct answer to a problem without knowing all the problem concepts. As a result, some attempts may not be an accurate representation of student knowledge.


\subsection{The Proposed Model}
\label{sec:proposed}
\begin{figure}[!ht]
    \centering
    \includegraphics[width=0.8\textwidth]{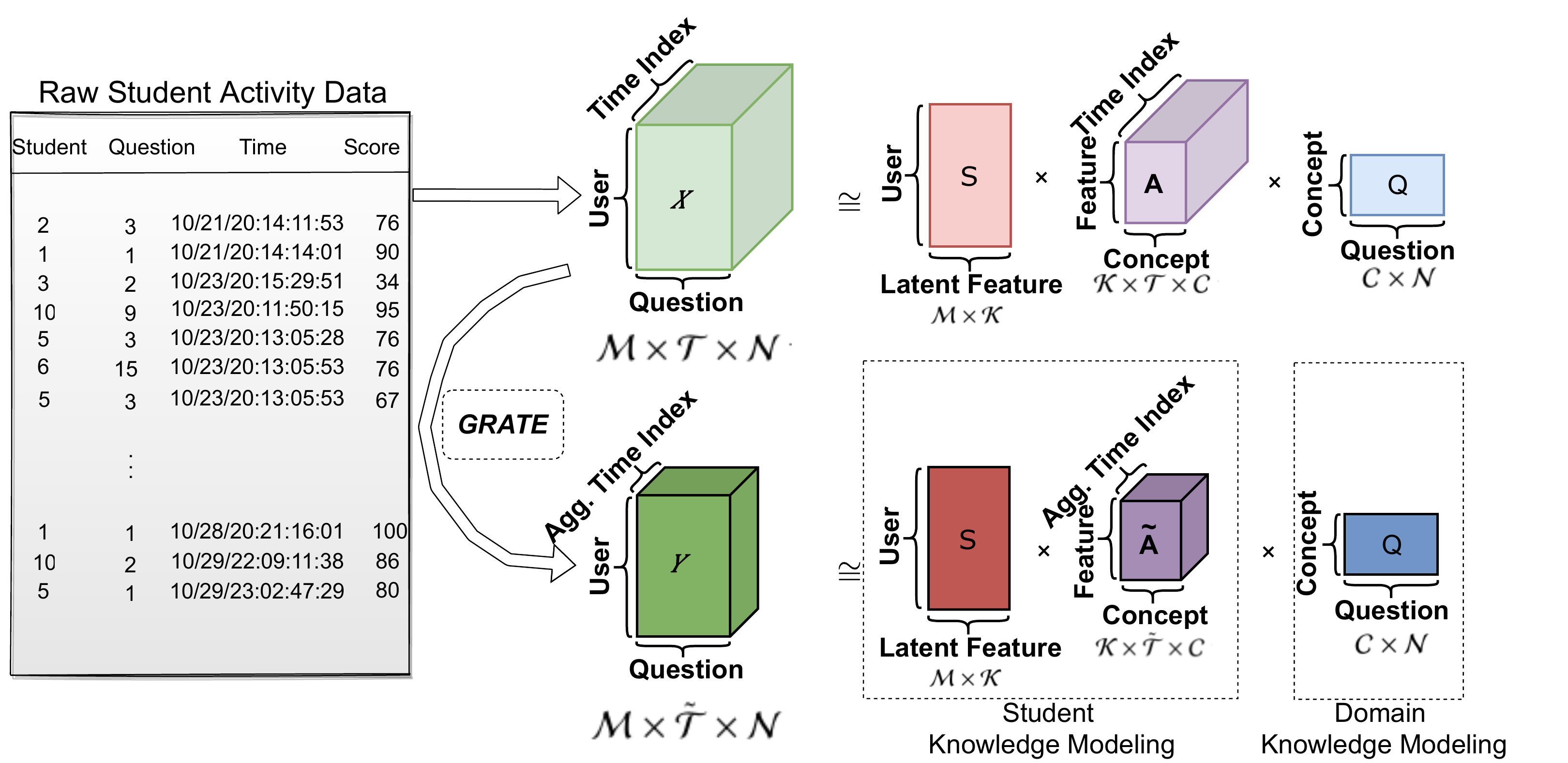}
    \caption{Knowledge Modeling via Granular Rank-Based Tensor Factorization.}
    \label{fig:rgbtf}
\end{figure}

\noindent\textbf{Tensor Factorization.} 
Following the (a) \textit{domain knowledge} and (b) \textit{student performance} assumptions above, we first model student interaction tensor $\mathbf{X}$ as a factorization of three lower dimensional representations: 
1) an $\mathcal{M} \times \mathcal{K}$ student latent feature matrix $S$, that represents particular student features (such as abilities and personalities) that are constant over time;
2) a $\mathcal{K} \times \mathcal{T} \times \mathcal{C}$ temporal dynamic knowledge tensor $\mathbf{A}$, that shows the knowledge of students with specific abilities in the course concepts as they attempt the problems; and 
3) a $\mathcal{C} \times \mathcal{N}$ matrix $Q$ serving as a mapping between problems and course concepts.
The upper tensor factorization in Figure~\ref{fig:rgbtf} represents this model.
According to our factorization, the resulting tensor from product $\mathbf{K} = S\mathbf{A}$ represents student knowledge in each concept at each attempt.
In addition to the above factors, we add a student-specific bias $b_{u}$, problem-specific difficulty $b_i$, and average score offset $\mu$.
Consequently, we can estimate students' performance at attempt ${t}$ as in the following, where $\sigma$ represents a standard probit or logit link function:
\begin{equation}
    \hat{x}_{u,t,i} = \sigma(\mathbf{s}_{u} \cdot A_t \cdot \mathbf{q}_{i} + b_{u} + b_i + \mu)
\end{equation}
Note that using $\sigma$ makes the model interpretation flexible: both as an estimation of a real-valued score between zero and one, and as the probability value of the binary success or failure in solving a problem (as in a classifier).
To learn the parameters $S$, $A$, $Q$, and the biases, we can minimize the following objective function:
\begin{equation}
    \begin{split}
    \mathcal{L}_0 &= \sum_{u,{t},i \in \Omega_{obs}} \left(x_{u,{t},i} - \hat{x}_{u,{t},i} \right)^{2} + \lambda_s \|s_u\|^2_2 + \lambda_a \sum_{{t}=1}^{\mathcal{{T}}} \|{A}_{{t}}\|_F^2 
    \end{split}
    \label{eq:loss0}
\end{equation}
in which the set $\Omega_{obs}$ consists of all non-missing values in $\mathbf{X}$. 
The last two terms are regularization constraints with important weights $\lambda_s$ and $\lambda_a$ to ensure the generalizability of the learned values.

\noindent\textbf{Adaptive Granularity-Based Aggregation.}
The student attempt tensor $\mathbf{X}$ is very sparse, since the students only interact with one problem during each attempt. 
Additionally, because of problem complexities, the observed attempts are noisy and unreliable (assumption (d) or \textit{attempt noisiness assumption}).
As a result, it is difficult to extract accurate and interpretable underlying structures from this tensor.
Moreover, equally relying on all attempts, whether they are noisy or not, results in imprecise knowledge states and poor performance predictions. 

To address this issue, we propose to automatically aggregate attempt slices that are deemed to be not informative enough with their neighbor attempts.
This will summarize the informative and non-informative attempts together, keeping only the important information and leaving out the noisy ones. 
To do this, we find a smaller aggregated tensor that summarizes our initial student attempt tensor, along with the attempts, with a maximal recovery of information or a minimal loss of information.

Inspired by~\cite{mlg2020_35}, 
we aim to find an aggregated 3-mode tensor $\mathbf{Y}$ of dimensions $\mathcal{M} \times \mathcal{\tilde{T}} \times \mathcal{N}$ with $\mathcal{\tilde{T}} \leq \mathcal{T}$, 
such that $\mathbf{Y}$ can accurately represent our initial tensor $\mathbf{X}$, measured by a goodness of fit or utility function $\mathbb{F}$.
To achieve this goal, we find an aggregation matrix $W$ that indicates which attempts should be integrated into their neighboring attempts. 
In other words, we find $W$ such that aggregating the $X$ tensor using it maximizes the utility function $\mathbb{F}$:
\begin{equation}
    \max_{W} \mathbb{F}(\mathbf{X}\times_{2} W)
    \label{prob:agg}
\end{equation}
where $\times_{2}$ denotes the aggregation operation on the tensor's second mode (attempt mode), and $W(i, j) = 1$ if slice $i$ in tensor $\mathbf{X}$ is aggregated into slice $j$ in the resulting tensor, otherwise $W(i, j) = 0$. 
For the utility function $\mathbb{F}$, we use missing value prediction accuracy.
As a result, we summarize the input tensor $\mathbf{X}$ using the aggregation matrix $W$ such that the new tensor $\mathbf{Y}$ provides us with the highest student performance prediction accuracy.
For instance, consider the tensor $\mathbf{X}$ of dimension $5 \times 6 \times 5$ with the optimal aggregate shape of $\mathbf{Y}$ is $5 \times 2 \times 5$. In this case,
$W$ is of size $2 \times 6$ such as:

\begin{equation}
W = 
\begin{bmatrix}
1 & 1 & 0 & 0 & 0 & 0\\
0 & 0 & 1 & 1 & 1 & 1
\end{bmatrix}
\end{equation}
where the first two slices of $\mathbf{X}$ will be aggregated as the first slice of $\mathbf{Y}$, and the last four will form the second. 

Since student knowledge states typically exhibit a Markovian property~\cite{zhao2020modeling}, we follow a greedy aggregation of slices, similar to the IceBreaker algorithm~\cite{mlg2020_35}, to try to aggregate the subsequent attempts.
This way, the aggregation matrix $W$ will have ones only close to the diagonal and zeros in all other places. 
This kind of aggregation also reduces the time complexity of solving Equation~\ref{prob:agg} from $O(2^{\mathcal{T}})$ to linear time complexity.

For the aggregation operation $\times_{2}$, we fill the unobserved values in the previous ($t-1$) attempt slice with the observed values in the current ($t$) attempt slice.
Also, we replace the observed values in the previous attempt with the values in the current attempt if they are observed.
This means that the aggregated attempt contains part of the observations from both the previous and current attempts. 
This is based on an intuitive assumption that the later attempts on a question are more informative than the earlier ones.
For instance, if students slipped once on the same question, they are less likely to slip again. 
Therefore, we use the current attempt values if they are available; otherwise, we use the less informative observations from the previous attempt.
The mathematical representation of our aggregation operation is summarized below~\footnote{Please note that our aggregation operation is different from in~\cite{mlg2020_35} in order to match our KT application.}.
\begin{equation}
    y_{u,\tilde{t-1},i} =
    \begin{cases}
        x_{u,t-1,i} \ &\text{if $x_{u,t-1,i}$ is observed, but $x_{u,t,i}$ is not observed} \\
        x_{u,t,i} \ &\text{if $x_{u,t,i}$ is observed}
    \end{cases}
\end{equation}

By having such an aggregation, we expect to rule out the noisy attempts that are observed due to the complexity of problems being solved.
To learn the parameters of this aggregated model, we update the loss in Equation~\ref{eq:loss0} in two ways to achieve the loss function in Equation~\ref{eq:loss_1}.
First, we use the new aggregated tensor $\mathbf{Y}$ instead of the input tensor $\mathbf{X}$\footnote{Please note that the aggregation and parameter estimation are done simultaneously in the algorithm, rather than by first learning a fixed $\mathbf{Y}$ and then by learning the parameters (Section~\ref{sec:alg}).}.
Second, to reduce the noise even further, we use the weighted least square error with confidence $\omega_{u,\tilde{t},i}$ as our objective function. 
We set $\omega_{u,\tilde{t},i}$ to be the number of student $u$'s trials on question $i$ by attempt $\tilde{t}$, to represent the confidence in that attempt's data.
This means that the more the student attempts a question, the more confidence we have in our observation of student performance in it.
\begin{equation}
    \begin{split}
    \mathcal{L}_1 &= \sum_{u,\tilde{t},i \in \Omega_{obs}} \omega_{u,\tilde{t},i} \cdot \left(y_{u,\tilde{t},i} - \hat{y}_{u,\tilde{t},i} \right)^{2} + \lambda_s \|S\|^2_F + \lambda_a \sum_{\tilde{t}=1}^{\mathcal{\tilde{T}}} \|\tilde{A}_{\tilde{t}}\|_F^2 
    \end{split}
    \label{eq:loss_1}
\end{equation}

\noindent\textbf{Rank-Based Knowledge Increase Constraint.} 
Finally, while we assume that student knowledge increases as a result of solving more problems (\textit{student learning} assumption (c)), the model so far does not follow such an assumption.
To address this, we add a rank-based constraint to the tensor factorization model, inspired by~\cite{doan2019rank}.
In particular, we would like to make sure that student knowledge increases in the problem-related concepts after the student interacts with a specific problem. 
That is $\mathbold{s}_t \cdot A_{t+1} \geq \mathbold{s}_t \cdot A_t$ or $\mathbold{k}_{u,t+1} \geq \mathbold{k}_{u,t}$.
Please note that, unlike~\cite{doan2019rank} where this assumption is imposed on predicted performance values, our assumption is imposed on student knowledge. 
This new constraint helps us to not only rely on the sparse and noisy performance observations in the data, but also to exploit the predicted (and unobserved) student performances.
This means that we can use the estimated student scores $\hat{x}_{u,t+1,i}$ for alleviating the noisiness and sparsity of the data tensor and prevent over-fitting to it. 
Finally, to add this assumption, we minimize the rank-based constraint in Equation~\ref{eq:loss_2}. 
This constraint will lead to a smooth and global knowledge increase in students, while allowing for small decreases, in case the student forgets some concepts. 
\begin{equation}
L_2 = \log \left( \sigma \left( \mathbold{s}_{u} \cdot \tilde{A}_{\tilde{t}+1} -  \mathbold{s}_{u} \cdot \tilde{A}_{\tilde{t}} \right) \right)
    \label{eq:loss_2}
\end{equation}

\noindent\textbf{The Full Model.}
Eventually, we put together all parts of the model to factorize the aggregated performance tensor with a rank-based constraint and learn the model parameters, including the student latent feature matrix, the knowledge tensor, and the problem concepts.
To this end, we optimize the objective function in Equation~\ref{eq:minimization}, which is a weighted combination of previously defined loss functions. 
This way, the adaptive granularity aggregation and the rank-based knowledge increase constraint cooperate to both resolve the sparsity problem and remove the noisy observations, which leads to a more accurate model.
Here, $\eta \in [0,1]$ is the trade-off parameter for the knowledge increase 
constraint versus the performance prediction fit. 
The outer summation can be replaced with implicit feedback sampling for some $j < \tilde{t}-1$ instead of the summation from $j=1$ to $j=\tilde{t}-1$. 

\begin{align}
    \mathop{\textbf{minimize}}_{\mathbf{s}_{u}, \tilde{A}_{\tilde{t}}, \mathbf{q}_i, b_u, b_i} &\mathcal{L} = \mathcal{L}_1 + \eta \sum_{j=1}^{\tilde{t}-1} \sum_{u, i} \log \left( \sigma \left( \mathbold{s}_{u} \cdot \tilde{A}_j -  \mathbold{s}_{u} \cdot \tilde{A}_{j+1} \right) \right) 
    \label{eq:minimization}
    \\
    \quad \text{subject to}&
    \begin{cases} 
    s_{u,i} \geq 0 &\forall u \in \{1,\cdots, \mathcal{M}\}, \forall i \in \{1, \cdots, \mathcal{K}\} \\
    q_{c,i} \geq 0  &\forall c \in \{1,\cdots, \mathcal{C}\}, \forall i \in \{1, \cdots, \mathcal{N}\} \\
    \sum_{c} q_{c,i} = 1.& \forall i \in \{1, \cdots, \mathcal{N}\} 
    \end{cases}
    \label{eq:constraints}
\end{align}

To ensure interpretable student modeling and proportional concept coverage on each problem, we enforce the constraints in Equation~\ref{eq:constraints}.
Specifically, the first constraint on $S$ ensures that student latent feature weights or student soft membership values in the latent features are non-negative.
The second and third constraints on $Q$ make sure that the concept contributions in each problem are non-negative and sum to one.
This way, we can interpret the $Q$ values as how much information on each concept is provided in each problem and can directly compare the problems with each other, according to their concepts.
We use a projected stochastic gradient descent on the final loss $\mathcal{L}$ to learn the parameters of the model.

\noindent\textbf{Our Intuition on Choice of the Model.} 
We have six main reasons for modeling the student knowledge tracing challenge as a tensor factorization problem.
First, unlike deep learning models, tensor factorization models do not require an abundance of data. 
Our model can work well with medium-sized datasets that have missing and noisy information.
Second, our tensor factorization design is perfect for discovering the underlying domain knowledge or concept structure.
The problem latent factors discovered by our model can be interpreted as the discovered concepts and the problems can be explained as weighted combinations of the discovered concepts.
In other words, problems are soft members of the concepts, and a concept's importance to each problem can be presented as the membership weight.
This is unlike KT models, such as PFA~\cite{pavlik2009performance}, that rely on a predefined domain knowledge model, or those like BKT~\cite{corbett1994knowledge} and iBKT~\cite{yudelson2013individualized} that model each concept independently.
Third, our proposed factorization is designed for personalized KT in students, following a collaborative-filtering~\cite{koren2009matrix,mnih2008probabilistic} type of structure in recommender systems.
Our model automatically attributes students to student latent features, which can be interpreted as student abilities and learning pace.
Similar to problems, students have soft memberships in these latent features and a separate set of personalized knowledge increase parameters are learned for each latent feature.
This contrasts with many existing deep learning based models\cite{piech2015deep, zhang2017dynamic, choi2020towards, ghosh2020context} that learn the same set of parameters for all students. 
Fourth, our choice of tensor factorization allows us to have a ``global'' trend on student knowledge increase.
Adding a constraint, similar to our rank-based constraint, is intrinsic to the tensor factorization models~\cite{xiong2010temporal,doan2019rank}. 
While the rank-based constraint is applied locally, it will result in a global increasing trend in student knowledge in the concepts that the student is interacting with.
This is similar to IRT~\cite{reckase2009multidimensional} and FTDF~\cite{sahebi2016tensor} models and is different from models such as DKT~\cite{piech2015deep} and DKVMN~\cite{zhang2017dynamic} that cannot easily constraint student knowledge trends in a global way.
Fifth, unlike models such as BKT~\cite{corbett1994knowledge} that have a two-state (known vs. unknown) knowledge state for students, our model can estimate \textit{how much} a student knows about each concept.
Finally and most importantly, by using the tensor structure, we can measure the importance of each attempt in student sequences and easily aggregate the less informative ones into previous attempts. 

\subsection{Algorithm and Implementation Details}
\label{sec:alg}
The pseudo-code of our GRATE model is described in Algorithm \ref{alg:rgbtf}~\footnote{The source code is provided at: \url{https://github.com/persai-lab/UMAP2021-GRATE}}.
In lines 1-2 of the algorithm, we predict students' performance at attempt 2. 
Lines 3-20 corresponds to the online training of the rest of the attempts.
For lines 2 and 19, we leverage the projected stochastic gradient descent to alternately update parameters ($\mathbf{s}_{u}, \tilde{A}_{\tilde{t}}, \mathbf{q}_i, b_u,$ and $b_i$) to solve the minimization problem (\ref{eq:minimization}). 
Moreover, to ensure $s_{u,i} \geq 0 \ \forall u,i$ and $q_{c,i} \geq 0\ \forall c,i$, the updated $s_{u,i}$ and $q_{c,i}$ are projected to the feasible set by clipping the values into $[0,1]$. To ensure $\sum_{c} q_{c,i} = 1, \forall i \in \{1, \cdots, \mathcal{N}\}$, we normalize the vector $\mathbf{q}_{i}$. 
We start with a large learning rate for SGD and reduce it by receiving new observation data. 
At line $12$, we use the online training mechanism in which we only train on new data points when new observation data are added. 
The algorithm's time complexity is linear with respect to the number of observations. Since each student could only work on a single question at each time index and the aggregation does not increase the number of observations, the time complexity is $O(\mathcal{M}\mathcal{T})$. 

\begin{algorithm}[!ht]
\footnotesize
\KwIn{Observed students' interaction records $\Delta_{obs}$, including training students' historical interaction records and targeting students' first interaction records;}
$\Omega_{obs} = \{x_{u,1,i} \in \Delta_{obs}, \forall u,i \} \bigcup \{x_{u,2,i} \in \Delta_{obs}, \forall u,i\}$ \\
Solve the problem (\ref{eq:minimization}) with constraints (\ref{eq:constraints}) on $\Omega_{obs}$. \\
$\mathcal{Y}_{prev} = \{x_{u,1,i} \in \Delta_{obs}, \forall u,i \}$\\
Create a tensor $\mathbold{Y}$ with size $\mathcal{M}\times 1\times \mathcal{N}$ and fill $\mathcal{Y}_{prev}$ into it. \\
\For{each time index $2 \leq t < T$}{
    Fill target students' new records $x_{u,t,i}$ into $\Delta_{obs}$\\
    $\mathcal{X}_{t} = \{x_{u,t,i} \in \Delta_{obs}, \forall u,i\}$ \\
    $\mathcal{Y}_{curr} = \mathcal{Y}_{prev} \sqcup \mathcal{X}_{t}$ \Comment*[r]{The $\sqcup$ denotes aggregation operation.} 
    \uIf{$Utility(\mathcal{Y}_{curr}) > Utility(\mathcal{Y}_{prev})$}{
        $\mathcal{Y}_{prev} = \mathcal{Y}_{curr}$. \\
        Fill $\mathcal{Y}_{curr}$ into last slice of $\mathbold{Y}$.
    }
    \Else{
        $\mathcal{Y}_{prev} = \mathcal{X}_{t}$. \\
        Add a new slice into $\mathbold{Y}$, and fill $\mathcal{X}_{t}$ into this new slice.
    }
    $\mathcal{X}_{t+1} = \{x_{u,t+1,i} \in \Delta_{obs}, \forall u,i\}$ \\
    Add a new slice into $\mathbold{Y}$, and fill $\mathcal{X}_{t+1}$ into this new slice. \\
    $\Omega_{obs} = \{\text{non-missing}\ y_{u,\tilde{t},i}\ \text{in}\ \mathbold{Y} \} $ \\
    Solve the problem (\ref{eq:minimization}) with constraint (\ref{eq:constraints}) on $\Omega_{obs}$.\\
}
\caption{GRATE Knowledge Modeling.}
\label{alg:rgbtf}
\end{algorithm}

%% file: 5_experiments.tex
We conducted experiments on three datasets to evaluate: 1) the model's performance in the student performance prediction task, as compared to the baselines; 2) the effect of different model parts on its performance; 3) the aggregation effect on student knowledge states; and 4) the discovered latent concepts.

\subsection{Datasets}
Three representative real-world datasets are used in our experiments.
Table \ref{table:datasets} shows the descriptive statistics of these three datasets.
\textbf{MORF} \footnote{\url{https://educational-technology-collective.github.io/morf/}} is a framework for accessing and experimenting with MOOC data of over 77 unique courses\cite{andres2016replicating}. 
We use the data from one of these courses, ``Big Data in Education'', as our experimental dataset.
This course includes 10 complex assignments, each of which contain a set of questions that are related to each other.
As only the students' grades on the whole assignments are available, we view each assignment as a complex ``problem'' in our model that covers multiple learning concepts.
These assignments are published in sequential order, but students can have multiple attempts on each assignment at any time. Students' scores are normalized into $[0,1]$.
\textbf{CSIntro}\footnote{\url{http://www.machineteaching.tech/en/}} contains students' outcomes from 10 different Computer Science introductory courses collected during two semesters, using the Machine Teaching ITS~\cite{moraes2020designing}. 
For each problem, the students write a Python program to solve the proposed problem.
The problems are complex, as solving each problem requires an understanding of multiple programming concepts, such as variables, loops, and conditional statements.
Student codes are tested against several test cases. The grade given at each attempt is the percentage of correct test cases. 
\textbf{MasteryGrids}\footnote{\url{http://adapt2.sis.pitt.edu/wiki/Mastery_Grids_Interface}} is a dataset collected from students' interactions with the Mastery Grids interface~\cite{loboda2014mastery} to learn the Python programming language. 
It contains students' binary scores (success or failure) on the programming problems in the system. 
Similar to CSIntro dataset, each problems requires students to understand a variety of concepts.
Figure \ref{fig:example_questions} shows some examples of the questions.

\begin{table}[!ht]
\footnotesize
\caption{Statistics of 3 Datasets.}
\centering
\resizebox{\textwidth}{!}{
\begin{tabular}{|c|c|c|c|c|c|c|c|c|}
\hline
Dataset & \#Users & \#Attempts & \#Problems & \#Records & \begin{tabular}[c]{@{}l@{}}Median \\ \# Problem \\Per User\end{tabular}  & \begin{tabular}[c]{@{}l@{}}Median \\ \#Attempts \\Per User\end{tabular}   & \begin{tabular}[c]{@{}l@{}} Median \\ \#Users \\Per Problem\end{tabular}  & \begin{tabular}[c]{@{}l@{}}Median \\ \#Attempts \\Per Problem\end{tabular} \\
\hline
MORF &686 &25 &10 &11700 &8  &17  &643.5 &1282 \\ 
\hline
CSIntro &120 &50 &48 &2231 &6  &16.5 &17 &37.5 \\
\hline
MasteryGrids &382 &70 &30 &10357 &14 &26 &166.5 &322 \\
\hline
\end{tabular}
}
\label{table:datasets}
\end{table}


\subsection{Experimental Setup and Baselines}
We use a \textit{5-fold nested student-stratified cross validation}
for our experiments. 
We shuffled and split the students randomly into five groups and ran the experiments for five rounds. 
At each round, we selected one of the five student groups as the test group and the rest (four groups) as the training group.
For test students, we started by predicting their performance at second attempt, given their first attempt. Then, we continued by predicting their performance at the next attempts, one by one, using their previous attempts and the training students' data.
We reported the average performance over five folds of testing data, as well as its calculated $95\%$ confidence interval. 
In order to see how the proposed model performed over time, we performed online training and testing, in which we predict the test data attempt by attempt.
For hyper-parameter tuning on our method and all baselines, we perform grid search over $25\%$ of the training data that is separated for validation. 
For GRATE, we grid searched over $\mathcal{K}, \mathcal{C}\in \{3,5,\cdots,17,19\}$, $\lambda_s, \lambda_a \in \{0.001, 0.005, 0.01, 0.05, 0.1\}$, and $\eta \in \{0.001, 0.01, 0.1, 0.2, 0.3\}$.
The best hyper-parameters are listed in Table~\ref{table:parameters}.

\begin{table}[!ht]
\footnotesize
\caption{Hyper-parameters of GRATE on three datasets.}
\centering
\resizebox{0.55\textwidth}{!}{
\begin{tabular}{|c|c|c|c|c|c|c|c|c|}
\hline
Dataset & $\mathcal{M}$ & $\mathcal{T}$ & $\mathcal{N}$ & $\mathcal{K}$ & $\mathcal{C}$ & $\lambda_s$ & $\lambda_a$ &$\eta$ \\
\hline
MORF &686 &25 &10 &3 &9 &0.001 &0.001 &0.1 \\ 
\hline
CSIntro &120 &50 &48 &7 &9 &0 &0.01 &0.2 \\ 
\hline
MasteryGrids &382 &70 &30 &3 &9 &0 &0 &0.01 \\ 
\hline
\end{tabular}
}
\label{table:parameters}
\end{table}

To evaluate the proposed approach's performance, we compare it with six different baseline models that cover a range of approaches to the task of knowledge tracing, from hidden Markov models to tensor factorization and deep learning.
A more detailed comparison of our model with these baselines can be found in the last part of Section~\ref{sec:proposed}  (model intuition).
Here is a short description of each of the baseline methods. 
\begin{itemize}
\item\textbf{DKT}: uses recurrent neural networks to model student learning on one concept at a time~\cite{piech2015deep}. DKT does not have the student knowledge modeling and domain knowledge modeling components. The performance is typically not interpretable.
\item\textbf{DKVMN}: is a variant of memory-augmented neural networks that tracks student states on multiple concepts~\cite{zhang2017dynamic}. Our model and DKVMN share some similarities: 1) DKVMN also uses a dynamic matrix to model student's knowledge state; 2) a student's performance on a specific question is determined by the concept coverage of the question and the student's mastery level of each concept; and 3) the Markov property of the knowledge state is modeled. 
\item\textbf{iBKT}\footnote{iBKT code from \url{https://github.com/CAHLR/pyBKT}}: is an extended hidden Markov model variant of the standard Bayesian knowledge tracing, providing individualization on student priors, learning rate, guess, and slip parameters in students~\cite{johnsonscaling,yudelson2013individualized, pardos2011kt}. 
\item\textbf{PMF}: is a classic probabilistic matrix factorization method for estimating missing values in matrices~\cite{mnih2008probabilistic}. To apply it on sequential student data, 
we use the last attempt of each question by each time index to build the matrix, mask test students' next questions for prediction, and estimate student performance on masked questions using observed entries in the matrix.
\item\textbf{FDTF}\footnote{FTDF code from \url{https://github.com/persai-lab/Tensor-Factorization-EDM}}: is a knowledge tracing method based on tensor factorization that enforces a constraint to guarantee a monotonic increase in student knowledge as per attempt~\cite{sahebi2016tensor}. 
\item\textbf{BPTF}\footnote{BPTF code from \url{https://www.cs.cmu.edu/~lxiong/bptf/bptf.html}}: is a Bayesian tensor factorization method for recommender systems~\cite{xiong2010temporal}. Unlike our model, BPTF uses a Gaussian distribution to smooth the knowledge transitions.
\end{itemize}

\subsection{GRATE for Student Performance Prediction}

In our first set of experiments, we evaluated the proposed model's performance in the task of student performance prediction, as compared to the baselines.
The experimental results are shown in Table~\ref{table:kt_results}. 
Since in the MORF and CSIntro datasets, the scores are real numbers, we use the root mean squared error (RMSE) between the predicted and test scores for evaluation (a lower number is better).
In the MasteryGrids dataset, the student performance is recorded as a binary success or failure. 
As a result, we used the classification mode of our proposed model and evaluated the results in terms of area under the curve (AUC), in which the higher values represent better performance.

As we can see, our proposed method significantly outperforms most of the baselines. 
The aggregation and ranked constraint not only do not harm the results, but can improve them too.
Based on our results, iBKT lags behind other baselines and FTDF has the second-best performance in MORF and CSIntro, while DKT is better than FTDF in MasteryGrids. 
This could be because DKT was designed for predicting the probability of success in datasets with binary scores, while FTDF was originally designed for estimating the score values.
Deep learning-based methods (DKT and DKVMN) did not have the best performance among the baseline models.
For example, DKVMN performed poorly on the MORF dataset.
This could be because all the baseline approaches, especially the deep learning and Markov model ones, were designed for simple types of problems and are highly sensitive to student performance fluctuations, which result in less accurate performance predictions.
This could also be because of the medium size of our datasets with a fewer number of both problems and students. 
These results from deep learning models are in agreement with the conclusions in previous research that show that deep learning models are not always the best choice for KT tasks in terms of data efficiency, interpretability, and prediction accuracy~\cite{gervet2020deep,khajah2016deep,wilsonback,xiong2016going}.

\begin{table*}[!htbp]
\footnotesize
\caption{Student performance prediction results in each of the datasets. Average root mean square error (RMSE) and area under curve (AUC) over five folds are used to evaluate performance on datasets with numerical feedback and binary feedback, respectively. $*$ and $**$ denote significance levels of p-value$<0.1$ and $<0.05$ in GRATE's improvement over the baselines.}
\centering
\resizebox{0.55\textwidth}{!}{
\begin{tabular}{|c|c|c|c|}
\hline
& \multicolumn{1}{c|}{\ \ \ \ \ \ \ MORF\ \ \ \ \ \ \ } & \multicolumn{1}{c|}{\ \ \ \ \ \ CSIntro\ \ \ \ \ \ } & \multicolumn{1}{c|}{MasteryGrids} \\
\cline{2-4}
Method & \multicolumn{1}{c|}{RMSE} & \multicolumn{1}{c|}{RMSE} & \multicolumn{1}{c|}{AUC} \\
\hline
DKT & \multicolumn{1}{c|}{$0.2170^{*\ }$} &\multicolumn{1}{c|}{$0.4132^{**}$} &\multicolumn{1}{c|}{$0.6968^{\ \ }$} \\
DKVMN & \multicolumn{1}{c|}{$0.2608^{**}$} &\multicolumn{1}{c|}{$0.4202^{**}$} &\multicolumn{1}{c|}{$0.6881^{\ \ }$} \\
iBKT & \multicolumn{1}{c|}{$0.2470^{**}$} &\multicolumn{1}{c|}{$0.6215^{**}$} &\multicolumn{1}{c|}{$0.6048^{**}$} \\
PMF & \multicolumn{1}{c|}{$0.2118^{\ \ }$} &\multicolumn{1}{c|}{$0.3982^{*\ }$} &\multicolumn{1}{c|}{$0.5647^{**}$}\\
BPTF & \multicolumn{1}{c|}{$0.2235^{**}$} &\multicolumn{1}{c|}{$0.3796^{\ \ }$} &\multicolumn{1}{c|}{$0.6942^{\ \ }$}\\
FDTF & \multicolumn{1}{c|}{$0.2088^{\ \ }$} &\multicolumn{1}{c|}{$0.3752^{\ \ }$} &\multicolumn{1}{c|}{$0.6877^{\ \ }$} \\
GRATE & $\mathbf{0.2033}^{\ \ }$ &$\mathbf{0.3726}^{\ \ }$ &$\mathbf{0.7035}^{\ \ }$\\
\hline
\end{tabular}
}
\label{table:kt_results}
\end{table*}

\subsection{Ablation Study}
Next, we experimented to evaluate the effect of different model components on its performance.
For that, we conducted an ablation study, removing one component from the model each time: GRATE-W/O-Agg for GRATE without the adaptive granularity aggregation component and GRATE-W/O-Rank for GRATE without the rank-based component.
The results of comparing these two models with GRATE on the task of performance prediction are shown in Table~\ref{table:ablation_results}.
As indicated, the full model performed better than both GRATE-W/O-Agg and GRATE-W/O-Rank in all datasets. 
This shows that neither of these two components was dispensable for student knowledge modeling and proved our adaptive granularity aggregation and rank-based learning assumptions.

\begin{table*}[!htbp]
\footnotesize
\caption{Ablation study results. The average performance over five folds of data $\pm$ $95\%$ confidence intervals are reported.
}
\centering
\resizebox{0.65\textwidth}{!}{
\begin{tabular}{|c|c|c|c|}
\hline
& \multicolumn{1}{c|}{MORF} & \multicolumn{1}{c|}{CSIntro} & \multicolumn{1}{c|}{MasteryGrids} \\
\cline{2-4}
Methods & RMSE & RMSE &AUC \\ 
\hline
GRATE-W/O-Agg. & $0.2100\pm0.0140$ &$0.3809\pm0.0089$ & $0.6849\pm0.0219$ \\
GRATE-W/O-Rank & $0.2251\pm0.0109$ & $0.3879\pm0.0244$ & $0.6837\pm0.0297$ \\
GRATE & $\mathbf{0.2033\pm0.0081}$ & $\mathbf{0.3726\pm0.0116}$ & $\mathbf{0.7035\pm0.0240}$ \\
\hline
\end{tabular}
}
\label{table:ablation_results}
\end{table*}

\subsection{GRATE for Student Knowledge Modeling}
\begin{figure}[ht!]
    \centering
    \includegraphics[width=0.6\linewidth]{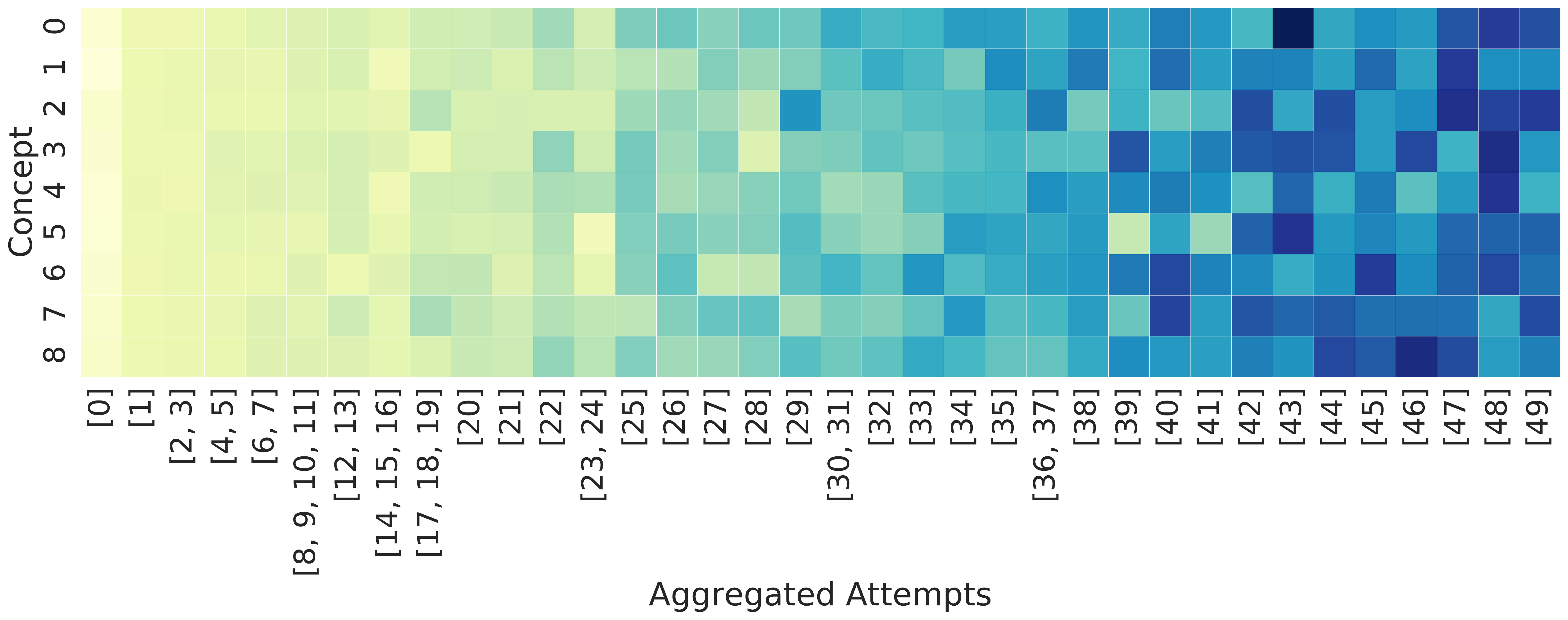}
    \caption{GRATE: Average knowledge transition over all students on CSIntro dataset. The darker and bluer cells represent higher knowledge values and the light and yellower ones represent lower knowledge values. (Best viewed in color).}
    \label{fig:rgbtf_avg_knowledge_growth}
\end{figure}

As mentioned in Section~\ref{sec:model}, GRATE can represent student knowledge over attempts via tensor $\mathbf{K} = S\mathbf{A}$.
In the following, we visualize student knowledge to study the interpretability of discovered knowledge values in students.
For easier visualization in two dimensions, we show the average student knowledge values, in each of the discovered latent concepts, over time.
The results for the CSIntro dataset are shown in Figure~\ref{fig:rgbtf_avg_knowledge_growth}.
The darker and bluer cells represent higher knowledge values, and the light and yellower ones represent lower knowledge values.
The X-axis shows the attempt numbers.
As we can see, students start with a low average knowledge on all concepts in their first attempt, and gradually gain knowledge as they try the problems. 
This increase is not strict, as in some attempts the students have a lower level of knowledge as compared to their previous attempt.
This is expected, as our rank-based knowledge increase constraint allows for occasional forgetting.
The learning rates between different concepts are different from each other.
Most importantly, we can see that some attempts are aggregated together and some are left as individual attempts. 
For example, attempts 4 and 5 are summarized together, and attempts 17, 18, and 19 are joined together. 

\begin{figure}[!ht]
    \centering
    \includegraphics[width=0.6\linewidth]{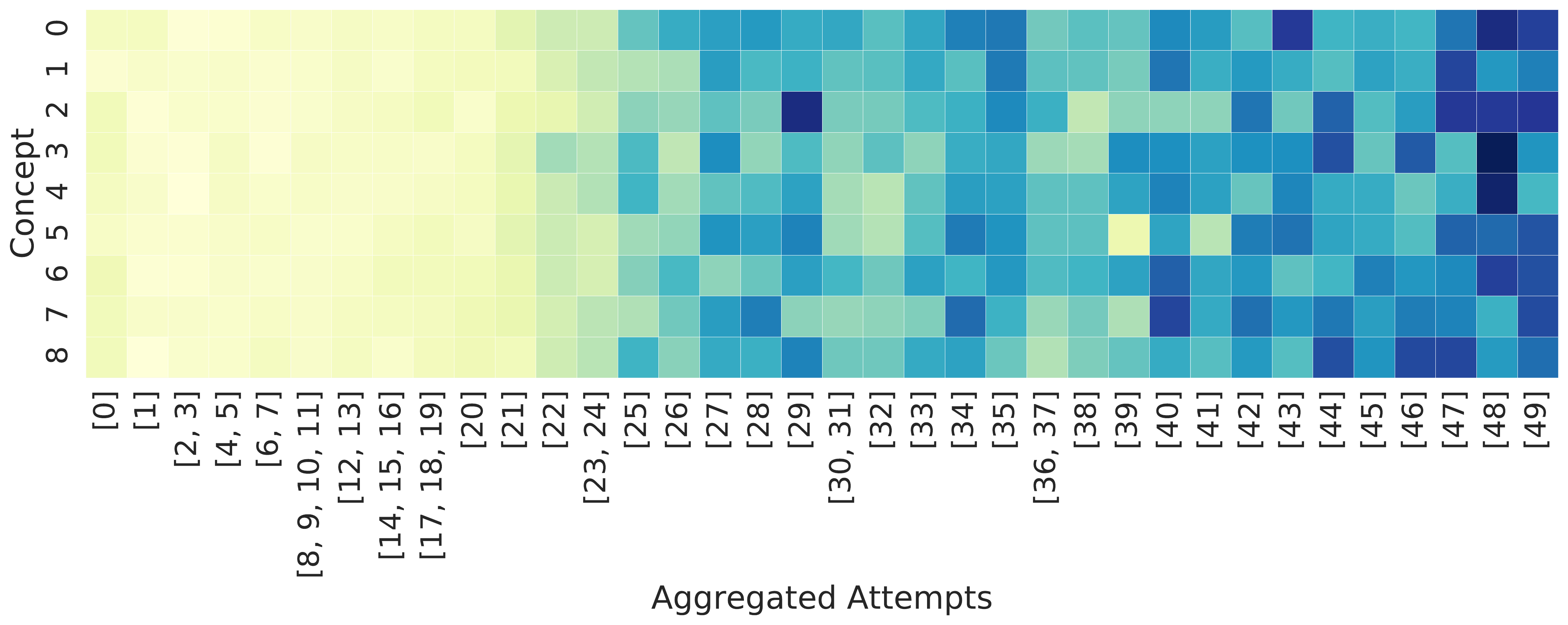}
    \caption{GRATE-W/O-Rank: Students' Average Knowledge State on CSIntro Dataset.}
    \label{fig:agtf_avg_knowledge_growth}
\end{figure}
\begin{figure}[!ht]
    \centering
    \includegraphics[width=0.9\linewidth]{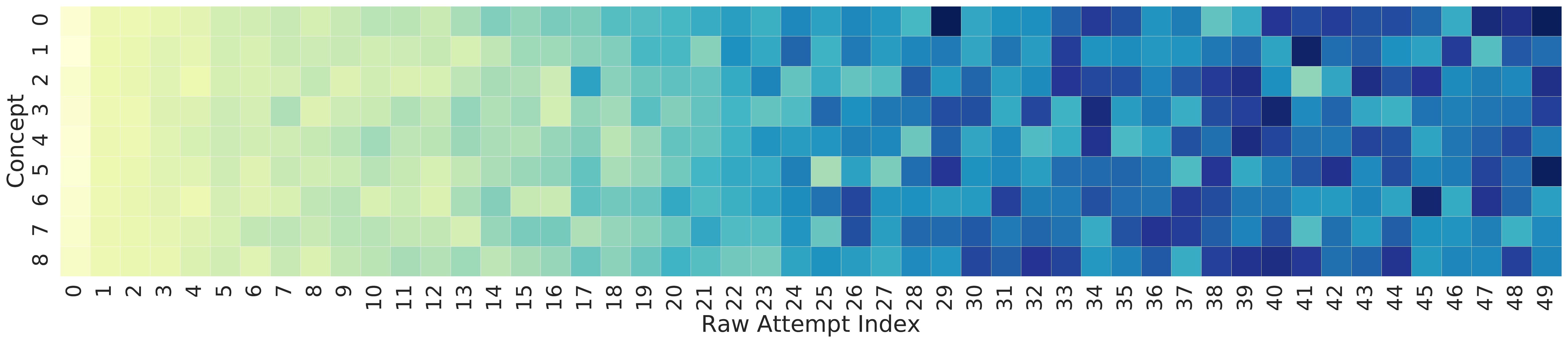}
    \caption{GRATE-W/O-Agg.: Students' Average Knowledge State on CSIntro Dataset.}
    \label{fig:rbtf_avg_knowledge_growth}
\end{figure}

To evaluate the rank-based constraint effect on student knowledge, we looked at the average student knowledge values calculated using the GRATE-W/O-Rank model shown in Figure~\ref{fig:agtf_avg_knowledge_growth}.
In this model, the attempts are still aggregated.
We can see that removing the rank-based constraint results in higher fluctuations in student knowledge values.
For example, students start with higher knowledge values in the first (0) attempt in Figure~\ref{fig:agtf_avg_knowledge_growth} only to lose it in the next attempt.
Contrasting this with Figure~\ref{fig:rgbtf_avg_knowledge_growth}, we see the effect of the knowledge increase constraint at the initial attempts.
As another example, student knowledge values on Concept 2 at Attempt 29 is much higher than their knowledge values at Attempts 28 and 30.
This value is much higher than the estimated knowledge values for Concept 2 at Attempt 29 in Figure~\ref{fig:rgbtf_avg_knowledge_growth}, when we have a rank-based constraint.

To assess why some of the attempts are aggregated together in GRATE and some are not, we look at the average student knowledge values in the GRATE-W/O-Agg model in Figure~\ref{fig:rbtf_avg_knowledge_growth}.
Here, although the rank-based constraint exists in the model, we see more fluctuations in the student knowledge values, as compared to the full GRATE model.
In particular, looking at the attempts that are aggregated together in Figure~\ref{fig:rgbtf_avg_knowledge_growth}, we see that high fluctuation attempts in Figure~\ref{fig:rbtf_avg_knowledge_growth} are aggregated together to create a smoother knowledge increase in Figure~\ref{fig:rgbtf_avg_knowledge_growth}.
For example, looking at Concept 2 at Attempts 17, 18, and 19 in Figure~\ref{fig:rbtf_avg_knowledge_growth}, we can see that the knowledge value drops from Attempt 17 (with value $k=0.30$) to 18 ($k=0.15$), just to increase again in Attempt 19 ($k=0.20$).
Similarly, looking at Concept 3 in the same attempts, we can see that the knowledge value decreases from Attempt 17 ($k=0.15$) to 18 ($k=0.13$), but then increases again in Attempt 19 ($k=0.22$).
However, in the full GRATE results in Figure~\ref{fig:rgbtf_avg_knowledge_growth}, these fluctuations are considered to be insignificant, and Attempts 17, 18, and 19 are aggregated together.
Another interesting observation is that the initial attempts are more aggregated together, as compared to the last attempts.
This could be because of the more confident knowledge estimations as more data is gathered about student learning in the final attempts.
Another potential reason is the low variability of knowledge among different students at the larger attempts, since a lesser number of students have longer sequences.

\subsection{GRATE for Domain Knowledge Modeling}
As shown in GRATE's model in Section~\ref{sec:proposed}, matrix $Q$ can represent latent concepts in different problems. In this experiment, we look at the $Q$-matrix discovered by GRATE in the Mastery Grids dataset as a case study, to make sense of the discovered latent concepts.
We chose Mastery Grids for this experiment, since each problem is assigned to a ``topic'' in this dataset.
We have not used these topics, or any form of content or annotated concepts, as an input to our model.
As a result, we use them as a standard to analyze the discovered matrix $Q$.

\begin{figure}[!ht]
    \centering
    \includegraphics[width=1.0\linewidth]{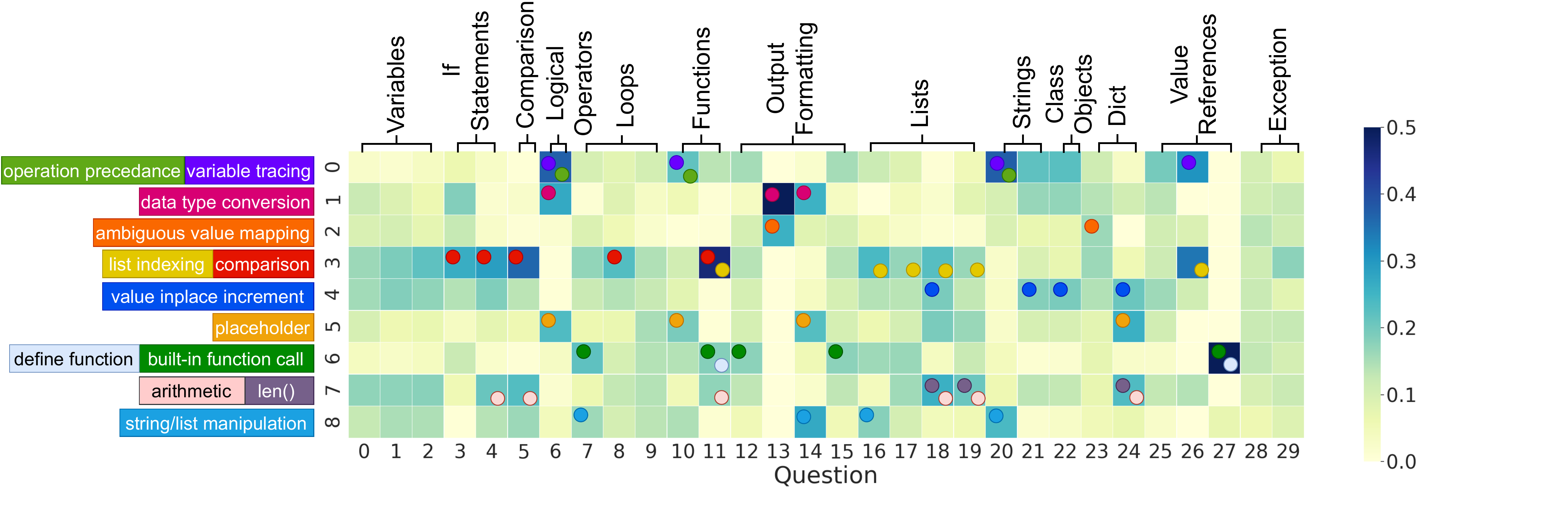}
    \caption{Matrix $Q$ discovered by GRATE for the Mastery Grids dataset. Each column represents a problem. Each row represents a latent concept. (Best viewed in color.)}
    \label{fig:mastery_grids_Q}
\end{figure}
Figure~\ref{fig:mastery_grids_Q} shows the matrix $Q$ discovered by GRATE for the Mastery Grids dataset.
The lower X-axis shows problem IDs, the upper X-axis shows problem topics from the dataset, and the Y-axis shows the discovered latent concepts.
Each cell is colored according to the latent concept weight in the problem.
The dark blue cells have a higher weight and the light yellow cells have a lower weight.
As the figure shows, 
many problems with the same topic have a high weight in similar latent concepts.
For example, Problems 3 and 4 that have the ``if statement" topic both have a high weight in latent concept 3 
and a low weight in latent concept 0. 
Or, Problems 16, 17, 18, and 19 that have the ``list indexing" topic all have a high weight in latent concept 3 
and a low weight in latent concepts 1 and 2. 
However, we can also see problems with different topics that have similar weights in latent concepts and problems with the same topic that have different important latent concepts.
For example, Problems 4 (with the ``if statement" topic) and 5 (with the ``comparison" topic) that both have a high weight in latent concepts 3 and 7. 

\begin{figure}[ht]
    \centering
    \includegraphics[width=0.75\linewidth]{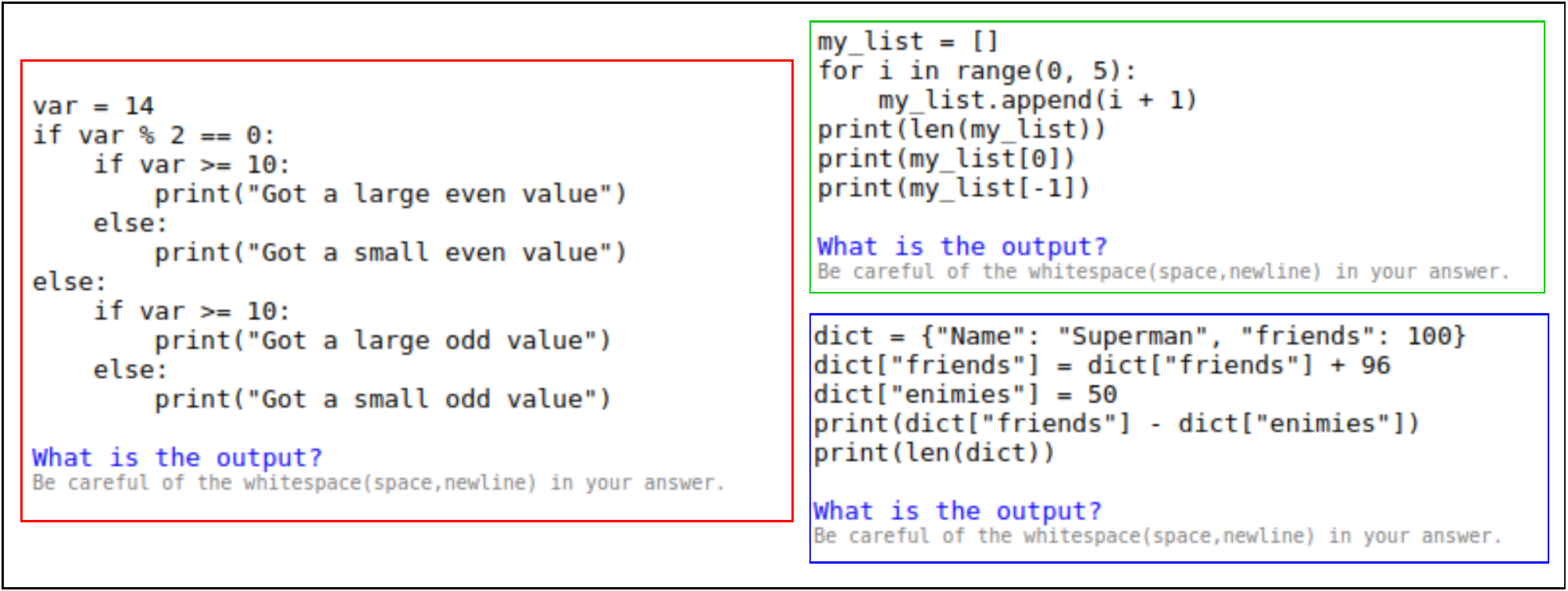}
    \caption{Problems 4 (left), 18 (top right), and 24 (bottom right) in MasteryGrids dataset.}
    \label{fig:example_questions}
\end{figure}
To better understand and analyze this phenomenon, we study and annotate each latent concept manually.
To this end, for each latent concept $q$, we compare the problems in which $q$ has a high weight and come up with a list of detailed concepts that are presented in them.
Then, we annotate the latent concept $q$ with the intersection of these concepts.
These annotations are color-coded and presented in the Y-axis of Figure~\ref{fig:mastery_grids_Q}, next to the latent concept numbers.
The problems that have a high weight of each annotated latent concept are marked by circles with the same color as the concept.
For example, latent concept 7 has a high weight in Problems $\{4, 5, 18, 19, 24\}$ that are from different topics.
In these problems, besides their major topic annotated in the dataset, two main concepts repeatedly appear: ``arithmetics" operations and ``len()" function.
For instance, we show Problems 4, 18, and 24 in Figure~\ref{fig:example_questions}.
We can see that in all of these three problems an arithmetic operation is necessary: Problem 4 computes the remainder of a division, Problem 18 has the addition operation, and Problem 24 has addition and subtraction operations. 
Meanwhile, both Problems 18 and 24 use the ``len()'' function to compute the length of a list and a dictionary, respectively.
We can see similar patterns in other problems with latent concept 7.
For this, we annotated Concept 7 with ``arithmetics" and ``len()".
We followed the same process for all latent concepts, as shown in Figure~\ref{fig:example_questions}.
This shows that our matrix $Q$ can unveil more detailed and complex structures in problems that go beyond simple labeling of them based on their ``topics''.
The discovered latent concepts not only represent conceptual similarities between problems, but also relate to student skills and performance in a detailed way.

%% file: 6_conclusions.tex
In this paper, we propose a solution to the knowledge tracing challenge for complex problem-solving. 
We argue that not all student attempts on complex problems are equally informative to estimate student knowledge and propose a granular rank-based tensor factorization (GRATE) model that adaptively aggregates consecutive student attempts for a more accurate estimate of student knowledge.
With experimenting on three real-world datasets we supported our argument by showing that:
(a) GRATE could perform better than state-of-the-art baselines in predicting student performance; 
(b) both aggregation and rank-based constraint are necessary for GRATE's superior performance; 
(c) GRATE improves student knowledge modeling by discovering and smoothing highly fluctuating and noisy attempts and implicitly detecting students' slip and guess; and 
(d) the latent concepts discovered by GRATE are meaningful and go beyond the labeled topics in the data.
Although GRATE's discovered concepts are latent and not directly human-readable, 
GRATE can guide teachers to determine which problems a student should practice to gain the required knowledge in these concepts.
The limitations of our study include the plausible but not analyzed assumptions, the medium size of our datasets, and not having access to the detailed expert annotations for labeling the latent concepts. Also, as the exactness of the discovered latent concepts depends on the accuracy of our model in predicting students' performance, the interpretability of the discovered matrix $Q$ can vary. 